\documentclass[showpacs,preprintnumbers,amsmath,amssymb]{revtex4}

\usepackage{graphicx}
\usepackage{dcolumn}
\usepackage{bm}
\begin{document}

\preprint{LA-UR 05-4348}

\title{Application of Vibration-Transit Theory of Liquid Dynamics to the Brillouin Peak Dispersion Curve}

\author{Duane C. Wallace}
\author{Giulia De Lorenzi-Venneri}
\author{Eric D. Chisolm}
\affiliation{Theoretical Division, Los Alamos National Laboratory, 
Los Alamos, New Mexico 87545}

\date{\today}

\begin{abstract}
The Brillouin peak appears in the dynamic structure factor $S(q,\omega)$, and 
the dispersion curve is the Brillouin peak frequency as function of $q$. The 
theoretical function underlying $S(q,\omega)$ is the density autocorrelation function $F(q,t)$. 
A broadly successful description of  time correlation functions is provided by 
mode coupling theory, which expresses $F(q,t)$ in terms of processes through 
which the density fluctuations decay. In contrast, vibration-transit (V-T) theory is a Hamiltonian formulation of
monatomic liquid dynamics in which the motion consists of vibrations within
a many-particle random valley, interspersed with nearly instantaneous transits
between such valleys.  Here, V-T theory is applied to $S(q,\omega)$. The theoretical 
vibrational contribution to $S(q,\omega)$ is the sum of independent scattering cross
sections from the normal vibrational modes, and contains no explicit reference to
decay processes. For a theoretical model of liquid Na, we show that the
vibrational contribution with no adjustable parameters gives an excellent 
account of the Brillouin peak dispersion curve, as compared to MD calculations
and to experimental data.
\end{abstract}

\pacs{05.20.Jj, 63.50.+x, 61.20.Lc, 61.12.Bt}
\keywords {Liquid Dynamics, Inelastic Neutron Scattering, Dispersion Relations, Mode Coupling Theory,
V-T Theory}
\maketitle

The purpose of this paper is to present a new theoretical explanation for
the Brillouin peak dispersion curve in a monatomic liquid.  The Brillouin 
peak represents inelastic neutron or photon scattering, and appears in
constant-$q$ graphs of the dynamic structure factor $S(q,\omega)$~\textit{vs}~$\omega$.
In monatomic liquids, the Brillouin peak is observable for $0 < q \lesssim q_{m}$, where
$q_{m}$ is at the maximum of the first peak in $S(q)$. At given $q$, the frequency
at the maximum of the Brillouin peak is $\Omega(q)$, and the dispersion
curve is $\Omega$~\textit{vs}~$q$. The new explanation results from the application of
vibration-transit (V-T) theory of liquid dynamics to the calculation
of $S(q,\omega)$. This theory was originally developed to explain the
equilibrium thermodynamic properties of monatomic liquids \cite{DWPRE56}. Now
that V-T theory has been shown to give a highly accurate account of
the thermodynamic properties of elemental liquids, with no adjustable
parameters \cite{DWPRE56,ChWJPCM01}, it is of interest to investigate its application to the
nonequilibrium problem of dynamic response.

The development of microscopic theories of liquid dynamics is described  in
the monographs of Boon and Yip \cite{BYbook}, Hansen and McDonald \cite{HMCDbook}, and
Balucani and Zoppi \cite{BZbook}. The theory most successful in accounting for time correlation
functions is mode coupling theory \cite{Golu75,GoSj92}. Here we shall limit our attention  to
the density autocorrelation function $F(q,t)$ and its Fourier transform $S(q,\omega)$. 
Mode coupling theory works with the generalized Langevin equation for $F(q,t)$,
and expresses the memory function in terms of processes through which
density fluctuations decay \cite{BYbook, HMCDbook, BZbook}.  In the viscoelastic approximation, the
memory function decays with a $q$-dependent relaxation time \cite{BYbook, HMCDbook, BZbook}. This
approximation provides a good fit to the combined experimental data \cite{CRPRL74} and
MD data \cite{ARPRL74,ARPRA74} for the Brillouin peak dispersion curve in liquid Rb \cite{CLRPP38} (see also Fig.~9.2
of \cite{HMCDbook}). Going beyond the viscoelastic approximation,
Bosse et al \cite{BGLPRA17a, BGLPRA17b} constructed a self-consistent theory for the
longitudinal and transverse current fluctuation spectra, each expressed in
terms of relaxation kernels approximated by decay integrals which couple
the longitudinal and transverse excitations. This theory is in good overall
agreement with extensive neutron scattering data and MD calculations
for Ar near its triple point \cite{BGLPRA17b}. The theory was developed further by
Sj\"ogren \cite{Sjog80, Sjog80b}, who separated the memory function into a binary
collision part, approximated with a Gaussian ansatz, and a more collective
tail represented by a mode coupling term. For liquid Rb, this theory
gives an ``almost quantitative'' agreement with results from neutron scattering experiments
\cite{CRPRL74} and MD calculations \cite{ARPRL74,ARPRA74}. More recently, inelastic x-ray scattering 
measurements have been done for the
light alkali metals Li \cite{R&c002} and Na \cite{SBRSb,Naexp}. In analyzing these data,
Scopigno et al \cite{SBRSb,Naexp,SRSVPRE02,SRSVPM02,SBRS,SBRSa}, include two or three decay
channels in the memory function, with each decay process represented by
adjustable $q$-dependent relaxation times and decay strengths. The resulting
fits to $S(q,\omega)$ are excellent, both for the experimental data and for MD
calculations \cite{SBRSb,Naexp,SRSVPRE02,SRSVPM02,SBRS,SBRSa}.

We shall now present a summary of V-T theory. From the outset, this 
theory was based on experimental data for the constant-volume entropy of
melting of the elements, and the constant-volume atomic-motion specific
heat of the elemental liquids at melt \cite{DWPRE56}. These data led us to the
conjecture that the potential energy surface consists entirely of intersecting
nearly-harmonic many-particle potential energy valleys; that these
valleys belong to two classes, the symmetric (microcrystalline) and
random classes; that the random valleys all have the same potential energy
surface in the thermodynamic limit; and that the random valleys are 
by far the most numerous, and therefore dominate the statistical mechanics 
of the liquid state. This conjecture has since been verified by computer
studies \cite{WCPRE59, CWPRE59}. The corresponding atomic motion consists of vibrations
within a single random valley, interspersed with motions across the
intersections between valleys, called transits. The vibrational contribution
dominates the thermodynamic properties. This contribution is modeled
by the motion of the system in a single harmonic random valley 
extended to infinity, and having no intersection with another valley. The
Hamiltonian then consists of a set of analytically tractable 
extended random valley Hamiltonians, plus corrections for transits and
for anharmonicity \cite{DWPRE56, DWbook2}. The role of transits in equilibrium is merely
to allow the system to move among all the random valleys, and hence
to exhibit the correct liquid entropy. The fact that transits contribute
little else to equilibrium thermodynamic properties is interpreted to indicate
that transits are nearly instantaneous \cite{DWPRE56, WChCPRE64}. Transits have a more
direct role in nonequilibrium properties, since transits are responsible 
for diffusive motion.

Since the vibrational contribution dominates the thermodynamic properties of the liquid, 
we expect it to be important in the nonequilibrium properties as well. 
This is the hypothesis to be tested. In fact, in the glass state,
where the system is frozen into a single random valley, the vibrational
contribution should be the \textit{only} contribution. We shall first check this
expectation in the glass, as its confirmation is required before we can
apply V-T theory to the liquid.

We have previously discussed  V-T theory for a general time correlation
function, and have derived the vibrational contribution to the dynamic response
functions \cite{WDCh05}. We work with a system containing $N$ atoms in a cubic volume,
with the motion governed by classical mechanics and by periodic boundary 
conditions. For motion in an extended random valley, the atomic positions
$\bm{r}_{K}(t)$ are written
\begin{equation} \label{eq1}
\bm{r}_{K}(t)=\bm{R}_{K}+\bm{u}_{K}(t),
\end{equation}
for $K=1, \dots,N$,where $\bm{R}_{K}$ is the equilibrium position of atom $K$,
and $\bm{u}_{K}(t)$ is its displacement. The vibrational contribution to $F(q,t)$ is $F_{vib}(q,t)$, and
in the one-mode approximation this is denoted $F_{1}(q,t)$. It is convenient to
separate the constant $F_{vib}(q,\infty)$, and write
\begin{equation} \label{eq2}
F_{1}(q,t)=F_{vib}(q,\infty)+\left[F_{1}(q,t)-F_{vib}(q,\infty) \right],
\end{equation}
where
\begin{equation} \label{eq3}
F_{vib}(q,\infty) =   \frac{1}{N} \left < \sum_{KL} e^{-i \bm{q}\cdot \bm{R}_{KL}} e^{-W_{K}(\bm{q})} e^{-W_{L}(\bm{q})}
\right >_{\bm{q}^{\ast}}, 
\end{equation}
\begin{equation}\label{eq4}
F_{1}(q,t)-F_{vib}(q,\infty)=
\frac {1}{N} \Bigg < \sum_{KL}e^{-i \bm{q}\cdot \bm{R}_{KL}}
\;e^{-W_{K}(\bm{q})} e^{-W_{L}(\bm{q})}
\;\left < \bm{q} \cdot \bm{u}_{K}(t)\;\bm{q} \cdot \bm{u}_{L}(0)\right >_{h}
                     \Bigg >_{\bm{q}^{\ast}} .
\end{equation}
Here $\bm{R}_{KL}=\bm{R}_{K}-\bm{R}_{L}$, $W_{K}(\bm{q})$ is the Debye-Waller factor for atom $K$,
$<\dots>_{h}$ is the harmonic average over the vibrational motion, and $<\dots>_{\bm{q}^{\ast}}$ is
the average over the star of $\bm{q}$. The atomic displacements are analyzed into
normal vibrational modes labeled $\lambda$, for $\lambda=1, \dots,3N$, each mode having frequency 
$\omega_{\lambda}$ and eigenvector components $\bm{w}_{K,\lambda}$. Then $S_{vib}(q,\omega)$,
the Fourier transform of $F_{1}(q,t)$, is
\begin{equation} \label{eq5}
S_{vib}(q,\omega) = F_{vib}(q,\infty)\delta(\omega)+S_{1}(q,\omega),
\end{equation}
where
\begin{equation} \label{eq6}
S_{1}(q,\omega)=\frac{3kT}{2M}\frac{1}{3N}\sum_{\lambda}f_{\lambda}(q)[\delta(\omega+\omega_{\lambda})+
\delta(\omega-\omega_{\lambda})], \\
\end{equation}
\begin{equation} \label{eq7}
f_{\lambda}(q)=\frac{1}{\omega_{\lambda}^{2}}\left < \left|\sum_{K}e^{-i\bm{q}\cdot\bm{R}_{K}}e^{-W_{K}(\bm{q})}
\bm{q}\cdot\bm{w}_{K\lambda}\right|^{2}\right >_{\bm{q}^{\ast}}.\\
\end{equation}
As a sum over normal modes, the Debye-Waller factors are
\begin{equation} \label{eq8}
W_{K}(\bm{q})=\sum_{\lambda}\frac{kT(\bm{q}\cdot\bm{w}_{K\lambda})^{2}}{2M\omega_{\lambda}^{2}}.
\end{equation}

Eq.~(\ref{eq6}) for  $S_{1}(q,\omega)$ has an obvious physical meaning: it is the sum over all vibrational 
modes of single-mode scattering at momentum transfer $\hbar q$, with creation 
or annihilation of energy $\hbar \omega_{\lambda}$ in mode $\lambda$. If crystal phonon eigenvectors
are used in Eq.~(\ref{eq7}), and if the average over $\bm{q}^{\ast}$ is omitted, one recovers the crystal
results of Lovesey \cite{Lovbook}, or of Glyde \cite{Glybook}, for $S_{1}(\bm{q},\omega)$.

\begin{figure}[t]
\includegraphics[height=3.0in,width=3.0in]{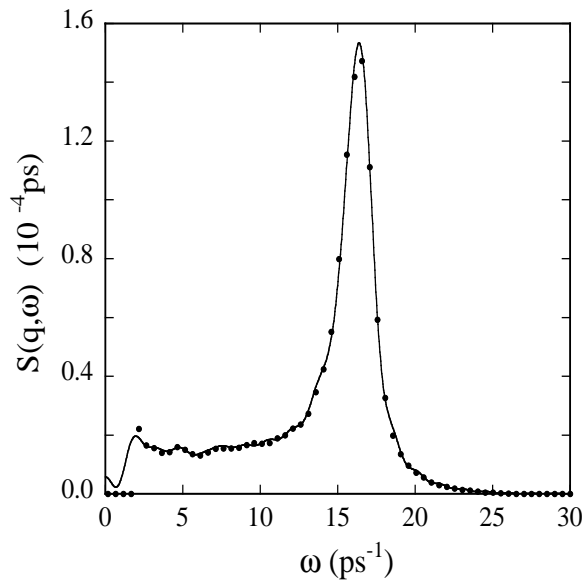}
\caption{\label{Sqw_T3} $S(q,\omega)$ for motion in a single random valley at $q=0.29711$ a$_{0}^{-1}$ and 24~K: 
full line is MD,
  points  are theory, Eq.~(\ref{eq6}).}
\end{figure}

The system we study is a model representing metallic Na, derived from 
electronic structure theory  (see e.g. \cite{DWbook2}). This model gives an excellent account
of the experimental phonon dispersion curves of bcc Na, and of the corresponding
thermal properties of the crystal (\cite{DWbook2}, Secs.~14, 17, 19). Through the spectrum
of the random valley vibrational mode frequencies $\omega_{\lambda}$, the model also accounts
for the experimental entropy of liquid Na (\cite{DWbook2}, Sec.~23 and Eq.~(24.2)). 
We shall use this model to evaluate the equations of V-T theory, and we shall
compare the results with MD calculations for the same model and with
experimental data on liquid Na. 
The same approach of comparing theory with MD was
successfully used by Glyde, Hansen, and Klein \cite{GHKPRB16} to evaluate the
contribution of anharmonicity to $S(\bm{q},\omega)$ for crystalline K.

Our first comparison is for the vibrational contribution alone, at a
representative $q$. For this purpose, the MD system was equilibrated at
24~K, where transits are frozen out and the system moves in a single random
valley. The MD result for $S(q,\omega)$ is shown by the solid line in Fig.~1.
Theory was evaluated from Eqs.~(\ref{eq6})-(\ref{eq8}), and is shown by the points in
Fig.~1. The $\delta(\omega)$ term in Eq.~(\ref{eq5}) expresses elastic scattering, and is not
indicated in the figure. The agreement between V-T theory and MD in Fig.~1 is excellent to the smallest
significant detail. Hence the same comparison for the liquid can be seen to reveal 
the effect of transits.

\begin{figure}[t]
\includegraphics[height=3.0in,width=3.0in]{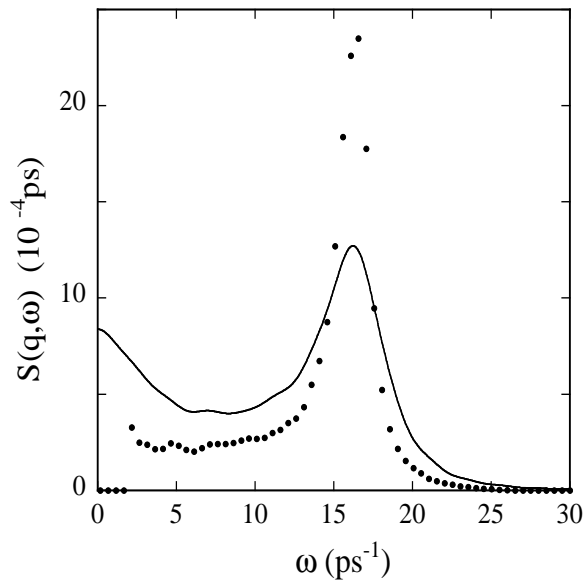}
\caption {\label{Sqw_T4}  $S(q,\omega)$ at $q=0.29711$ a$_{0}^{-1}$ and  395~K: line is MD for the liquid, 
points are the vibrational contribution alone, Eq.~(\ref{eq6}). What the vibrational contribution
gets right is the location of the Brillouin peak.}
\end{figure}

In previous work, Mazzacurati et al.\cite{Mazz96} (see also \cite{R&cPRL00}) have shown
that a vibrational analysis is in excellent agreement with MD calculations of $S(q,\omega)$
for a Lennard-Jones glass. We have done this comparison again, in Fig.~1, because 
we want to study the \textit{liquid}, and V-T theory requires confirmation
that the valley studied is random, and not some other symmetry. 

We shall now make the same comparison for the liquid state, at 395~K. 
For our system, this temperature is some 24~K above melting.  The vibrational
contribution to $S(q,\omega)$, again calculated from Eqs.~(\ref{eq6})-(\ref{eq8}), is shown by
the points in Fig.~2. For this contribution, the temperature dependence is
explicit in Eqs.~(\ref{eq6}) and (\ref{eq8}), and is dominated in Fig.~2 by the linear 
dependence in Eq.~(\ref{eq6}). On the other hand, the MD system at 395~K is
transiting rapidly among the random valleys. The MD result, the solid line
in Fig.~2, displays the well-known Rayleigh-Brillouin spectrum \cite{BYbook, HMCDbook, BZbook},
where the Rayleigh peak arises from the elastic term in Eq.~(\ref{eq5}).
What is missing from the V-T curve is the transit contribution. From the
comparison with the MD curve, which includes transits, we can see that the 
effect of transits is to broaden both the Rayleigh and the Brillouin peaks,
but not to shift them. That  these peaks are not shifted results from the
nearly instantaneous character of transits.

Three additional properties related to Fig.~2 are of interest. First, the
vibrational contribution alone gives an accurate account of the MD results for
the area of the Rayleigh peak, and for the area of the Brillouin peak. This
property of V-T theory is also understood from the nearly instantaneous 
character of transits. Second, the vibrational contribution alone also gives
around half of the total width of the Brillouin peak. And third, a
parameterized model for $S(q,\omega)$, in the spirit of Zwanzig's model for
the velocity autocorrelation function \cite{ZJCP79}, is obtained by multiplying the two
terms on the right side of Eq.~(\ref{eq2}) by the relaxation functions $e^{-\alpha_{1} t}$
and $e^{-\alpha_{2} t}$, respectively. Since they represent the transit-induced decay
of $F(q,t)$, the relaxation rates  $\alpha_{1}$ and $\alpha_{2}$ are expected to be approximately
the mean transit rate in the liquid \cite{DW05}. For the example shown in Fig.~2, and also 
for the other $q$ we have studied, a good fit to the MD $S(q,\omega)$ is obtained
from this relaxation model \cite{DW05}, with $\alpha_{1}$ and $\alpha_{2}$ close to the known 
mean transit rate \cite{DWPRE58,ChCWPRE63}.

\begin{figure}[h]
\includegraphics[height=3.0in,width=3.0in]{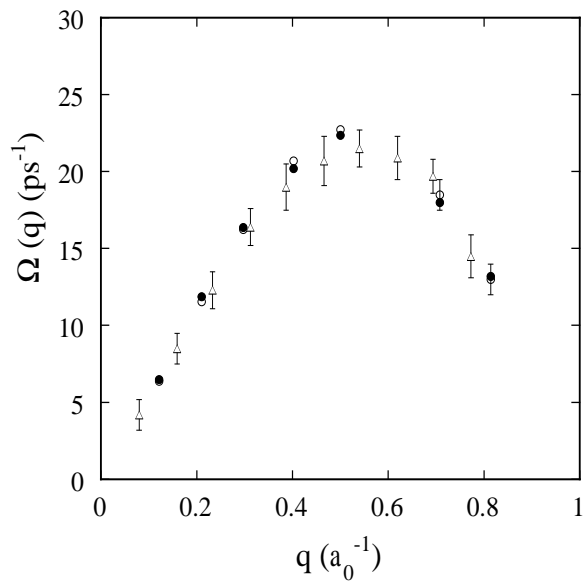}
\caption {\label{wq}  Brillouin peak dispersion curve for liquid sodium: circles are MD  at 395~K, points
   are the vibrational contribution from V-T theory at 395~K, and triangles are
   x-ray scattering data at 390~K \cite{SBRSb,Naexp}. }
\end{figure}

We shall now present the main result of this paper. Since transits
broaden the Brillouin peak, without shifting it, the Brillouin peak
dispersion curve is determined by the vibrational contribution alone.
In Fig.~3, $\Omega(q)$ is
graphed for the entire range of $q$ for which the Brillouin peak is observable in our system, comparing 
results from MD in the liquid at
395~K with the vibrational contribution alone at the same temperature. The
relative rms difference between the $\Omega_{MD}$ and $\Omega_{vib}$ points is 2$\%$, a difference on the
order of numerical errors. Experimental data for liquid Na at 390~K \cite{SBRSb,Naexp} are
also graphed, and our theoretical $\Omega_{vib}$ points fall within the experimental 
error bars. The relative rms difference of our seven $\Omega_{vib}$ points from a smooth curve
fitting the experimental points is 5$\%$.

In summary: (a) the vibrational contribution to $S(q,\omega)$ for a monatomic
liquid is the sum of independent inelastic scattering cross sections from
each of the random valley normal vibrational modes; (b) in contrast to
mode coupling theory, this formulation of $S(q,\omega)$ is not based on
explicit decay processes; and (c) for liquid Na, comparison with MD
calculations and with experimental data shows that the vibrational 
contribution accurately accounts for the location of the Brillouin peak,
and accounts for half the width of the Brillouin peak as well. 
Our results are consistent with the findings of Sciortino and Tartaglia \cite{ST97}
for water, that the vibrational motion is quite harmonic, and that the fast
decay of correlation functions is due to mode dephasing. Here, the dephasing
gives rise to the natural width of the vibrational contribution to
the Brillouin peak, as shown in Figs.~1 and 2.

   The authors gratefully acknowledge Tullio Scopigno and R. M. Yulmetyev for providing
   the original experimental data, Brad Clements for collaboration, and 
   Francesco Venneri for technical 
   advice. This work was supported by the US DOE through contract W-7405-ENG-36.

\bibliography{VTTref}

\begin{thebibliography}{37}
\expandafter\ifx\csname natexlab\endcsname\relax\def\natexlab#1{#1}\fi
\expandafter\ifx\csname bibnamefont\endcsname\relax
  \def\bibnamefont#1{#1}\fi
\expandafter\ifx\csname bibfnamefont\endcsname\relax
  \def\bibfnamefont#1{#1}\fi
\expandafter\ifx\csname citenamefont\endcsname\relax
  \def\citenamefont#1{#1}\fi
\expandafter\ifx\csname url\endcsname\relax
  \def\url#1{\texttt{#1}}\fi
\expandafter\ifx\csname urlprefix\endcsname\relax\def\urlprefix{URL }\fi
\providecommand{\bibinfo}[2]{#2}
\providecommand{\eprint}[2][]{\url{#2}}

\bibitem[{\citenamefont{Wallace}(1997)}]{DWPRE56}
\bibinfo{author}{\bibfnamefont{D.~C.} \bibnamefont{Wallace}},
  \bibinfo{journal}{Phys.\ Rev.\ E} \textbf{\bibinfo{volume}{56}},
  \bibinfo{pages}{4179} (\bibinfo{year}{1997}).

\bibitem[{\citenamefont{Chisolm and Wallace}(2001)}]{ChWJPCM01}
\bibinfo{author}{\bibfnamefont{E.~D.} \bibnamefont{Chisolm}} \bibnamefont{and}
  \bibinfo{author}{\bibfnamefont{D.~C.} \bibnamefont{Wallace}},
  \bibinfo{journal}{J.\ Phys.:\ Condens. Matter} \textbf{\bibinfo{volume}{13}},
  \bibinfo{pages}{R739} (\bibinfo{year}{2001}).

\bibitem[{\citenamefont{Boon and Yip}(1980)}]{BYbook}
\bibinfo{author}{\bibfnamefont{J.~P.} \bibnamefont{Boon}} \bibnamefont{and}
  \bibinfo{author}{\bibfnamefont{S.}~\bibnamefont{Yip}},
  \emph{\bibinfo{title}{Molecular Hydrodynamics}}
  (\bibinfo{publisher}{McGraw-Hill, New York}, \bibinfo{year}{1980}).

\bibitem[{\citenamefont{Hansen and McDonald}(1986)}]{HMCDbook}
\bibinfo{author}{\bibfnamefont{J.~P.} \bibnamefont{Hansen}} \bibnamefont{and}
  \bibinfo{author}{\bibfnamefont{I.~R.} \bibnamefont{McDonald}},
  \emph{\bibinfo{title}{Theory of Simple Liquids}}
  (\bibinfo{publisher}{Academic, New York}, \bibinfo{year}{1986}),
  \bibinfo{edition}{2nd} ed.

\bibitem[{\citenamefont{Balucani and Zoppi}(1994)}]{BZbook}
\bibinfo{author}{\bibfnamefont{U.}~\bibnamefont{Balucani}} \bibnamefont{and}
  \bibinfo{author}{\bibfnamefont{M.}~\bibnamefont{Zoppi}},
  \emph{\bibinfo{title}{Dynamics of the Liquid State}}
  (\bibinfo{publisher}{Clarendon Press, Oxford}, \bibinfo{year}{1994}),
  \bibinfo{edition}{2nd} ed.

\bibitem[{\citenamefont{G{\"o}tze and L{\"u}cke}(1975)}]{Golu75}
\bibinfo{author}{\bibfnamefont{W.}~\bibnamefont{G{\"o}tze}} \bibnamefont{and}
  \bibinfo{author}{\bibfnamefont{M.}~\bibnamefont{L{\"u}cke}},
  \bibinfo{journal}{Phys.\ Rev.\ A} \textbf{\bibinfo{volume}{11}},
  \bibinfo{pages}{2173} (\bibinfo{year}{1975}).

\bibitem[{\citenamefont{G{\"o}tze and Sj{\"o}gren}(1992)}]{GoSj92}
\bibinfo{author}{\bibfnamefont{W.}~\bibnamefont{G{\"o}tze}} \bibnamefont{and}
  \bibinfo{author}{\bibfnamefont{L.}~\bibnamefont{Sj{\"o}gren}},
  \bibinfo{journal}{Rep.\ Prog.\ Phys.} \textbf{\bibinfo{volume}{55}},
  \bibinfo{pages}{241} (\bibinfo{year}{1992}).

\bibitem[{\citenamefont{Copley and Rowe}(1974)}]{CRPRL74}
\bibinfo{author}{\bibfnamefont{J.~R.~D.} \bibnamefont{Copley}}
  \bibnamefont{and} \bibinfo{author}{\bibfnamefont{J.~M.} \bibnamefont{Rowe}},
  \bibinfo{journal}{Phys.\ Rev.\ Lett.} \textbf{\bibinfo{volume}{32}},
  \bibinfo{pages}{49} (\bibinfo{year}{1974}).

\bibitem[{\citenamefont{Rahman}(1974{\natexlab{a}})}]{ARPRL74}
\bibinfo{author}{\bibfnamefont{A.}~\bibnamefont{Rahman}},
  \bibinfo{journal}{Phys.\ Rev.\ Lett.} \textbf{\bibinfo{volume}{32}},
  \bibinfo{pages}{52} (\bibinfo{year}{1974}{\natexlab{a}}).

\bibitem[{\citenamefont{Rahman}(1974{\natexlab{b}})}]{ARPRA74}
\bibinfo{author}{\bibfnamefont{A.}~\bibnamefont{Rahman}},
  \bibinfo{journal}{Phys.\ Rev.\ A} \textbf{\bibinfo{volume}{9}},
  \bibinfo{pages}{1667} (\bibinfo{year}{1974}{\natexlab{b}}).

\bibitem[{\citenamefont{Copley and Lovesey}(1975)}]{CLRPP38}
\bibinfo{author}{\bibfnamefont{J.~R.~D.} \bibnamefont{Copley}}
  \bibnamefont{and} \bibinfo{author}{\bibfnamefont{S.~W.}
  \bibnamefont{Lovesey}}, \bibinfo{journal}{Rep.\ Prog.\ Phys.}
  \textbf{\bibinfo{volume}{38}}, \bibinfo{pages}{461} (\bibinfo{year}{1975}).

\bibitem[{\citenamefont{Bosse et~al.}(1978{\natexlab{a}})\citenamefont{Bosse,
  G{\"o}tze, and L{\"u}cke}}]{BGLPRA17a}
\bibinfo{author}{\bibfnamefont{J.}~\bibnamefont{Bosse}},
  \bibinfo{author}{\bibfnamefont{W.}~\bibnamefont{G{\"o}tze}},
  \bibnamefont{and}
  \bibinfo{author}{\bibfnamefont{M.}~\bibnamefont{L{\"u}cke}},
  \bibinfo{journal}{Phys.\ Rev.\ A} \textbf{\bibinfo{volume}{17}},
  \bibinfo{pages}{434} (\bibinfo{year}{1978}{\natexlab{a}}).

\bibitem[{\citenamefont{Bosse et~al.}(1978{\natexlab{b}})\citenamefont{Bosse,
  G{\"o}tze, and L{\"u}cke}}]{BGLPRA17b}
\bibinfo{author}{\bibfnamefont{J.}~\bibnamefont{Bosse}},
  \bibinfo{author}{\bibfnamefont{W.}~\bibnamefont{G{\"o}tze}},
  \bibnamefont{and}
  \bibinfo{author}{\bibfnamefont{M.}~\bibnamefont{L{\"u}cke}},
  \bibinfo{journal}{Phys.\ Rev.\ A} \textbf{\bibinfo{volume}{17}},
  \bibinfo{pages}{447} (\bibinfo{year}{1978}{\natexlab{b}}).

\bibitem[{\citenamefont{Sj{\"o}gren}(1980{\natexlab{a}})}]{Sjog80}
\bibinfo{author}{\bibfnamefont{L.}~\bibnamefont{Sj{\"o}gren}},
  \bibinfo{journal}{Phys.\ Rev.\ A} \textbf{\bibinfo{volume}{22}},
  \bibinfo{pages}{2866} (\bibinfo{year}{1980}{\natexlab{a}}).

\bibitem[{\citenamefont{Sj{\"o}gren}(1980{\natexlab{b}})}]{Sjog80b}
\bibinfo{author}{\bibfnamefont{L.}~\bibnamefont{Sj{\"o}gren}},
  \bibinfo{journal}{Phys.\ Rev.\ A} \textbf{\bibinfo{volume}{22}},
  \bibinfo{pages}{2883} (\bibinfo{year}{1980}{\natexlab{b}}).

\bibitem[{\citenamefont{Scopigno
  et~al.}(2000{\natexlab{a}})\citenamefont{Scopigno, Balucani, Cunsolo,
  Masciovecchio, Ruocco, Sette, and Verbeni}}]{R&c002}
\bibinfo{author}{\bibfnamefont{T.}~\bibnamefont{Scopigno}},
  \bibinfo{author}{\bibfnamefont{U.}~\bibnamefont{Balucani}},
  \bibinfo{author}{\bibfnamefont{A.}~\bibnamefont{Cunsolo}},
  \bibinfo{author}{\bibfnamefont{C.}~\bibnamefont{Masciovecchio}},
  \bibinfo{author}{\bibfnamefont{G.}~\bibnamefont{Ruocco}},
  \bibinfo{author}{\bibfnamefont{F.}~\bibnamefont{Sette}}, \bibnamefont{and}
  \bibinfo{author}{\bibfnamefont{R.}~\bibnamefont{Verbeni}},
  \bibinfo{journal}{Europhys.\ Lett.} \textbf{\bibinfo{volume}{50}},
  \bibinfo{pages}{189} (\bibinfo{year}{2000}{\natexlab{a}}).

\bibitem[{\citenamefont{Scopigno
  et~al.}(2002{\natexlab{a}})\citenamefont{Scopigno, Balucani, Ruocco, and
  Sette}}]{SBRSb}
\bibinfo{author}{\bibfnamefont{T.}~\bibnamefont{Scopigno}},
  \bibinfo{author}{\bibfnamefont{U.}~\bibnamefont{Balucani}},
  \bibinfo{author}{\bibfnamefont{G.}~\bibnamefont{Ruocco}}, \bibnamefont{and}
  \bibinfo{author}{\bibfnamefont{F.}~\bibnamefont{Sette}},
  \bibinfo{journal}{Phys.\ Rev.\ E} \textbf{\bibinfo{volume}{65}},
  \bibinfo{pages}{031205} (\bibinfo{year}{2002}{\natexlab{a}}).

\bibitem[{\citenamefont{Yulmetyev et~al.}(2003)\citenamefont{Yulmetyev,
  Mokshin, Scopigno, and H{\"a}nggi}}]{Naexp}
\bibinfo{author}{\bibfnamefont{R.~M.} \bibnamefont{Yulmetyev}},
  \bibinfo{author}{\bibfnamefont{A.~V.} \bibnamefont{Mokshin}},
  \bibinfo{author}{\bibfnamefont{T.}~\bibnamefont{Scopigno}}, \bibnamefont{and}
  \bibinfo{author}{\bibfnamefont{P.}~\bibnamefont{H{\"a}nggi}},
  \bibinfo{journal}{J.\ Phys.\ Condensed Matter} \textbf{\bibinfo{volume}{15}},
  \bibinfo{pages}{2235} (\bibinfo{year}{2003}).

\bibitem[{\citenamefont{Scopigno
  et~al.}(2002{\natexlab{b}})\citenamefont{Scopigno, Ruocco, Sette, and
  Viliani}}]{SRSVPRE02}
\bibinfo{author}{\bibfnamefont{T.}~\bibnamefont{Scopigno}},
  \bibinfo{author}{\bibfnamefont{G.}~\bibnamefont{Ruocco}},
  \bibinfo{author}{\bibfnamefont{F.}~\bibnamefont{Sette}}, \bibnamefont{and}
  \bibinfo{author}{\bibfnamefont{G.}~\bibnamefont{Viliani}},
  \bibinfo{journal}{Phys.\ Rev.\ E} \textbf{\bibinfo{volume}{66}},
  \bibinfo{pages}{031205} (\bibinfo{year}{2002}{\natexlab{b}}).

\bibitem[{\citenamefont{Scopigno
  et~al.}(2002{\natexlab{c}})\citenamefont{Scopigno, Ruocco, Sette, and
  Viliani}}]{SRSVPM02}
\bibinfo{author}{\bibfnamefont{T.}~\bibnamefont{Scopigno}},
  \bibinfo{author}{\bibfnamefont{G.}~\bibnamefont{Ruocco}},
  \bibinfo{author}{\bibfnamefont{F.}~\bibnamefont{Sette}}, \bibnamefont{and}
  \bibinfo{author}{\bibfnamefont{G.}~\bibnamefont{Viliani}},
  \bibinfo{journal}{Phil.\ Mag.\ B} \textbf{\bibinfo{volume}{82}},
  \bibinfo{pages}{233} (\bibinfo{year}{2002}{\natexlab{c}}).

\bibitem[{\citenamefont{Scopigno
  et~al.}(2000{\natexlab{b}})\citenamefont{Scopigno, Balucani, Ruocco, and
  Sette}}]{SBRS}
\bibinfo{author}{\bibfnamefont{T.}~\bibnamefont{Scopigno}},
  \bibinfo{author}{\bibfnamefont{U.}~\bibnamefont{Balucani}},
  \bibinfo{author}{\bibfnamefont{G.}~\bibnamefont{Ruocco}}, \bibnamefont{and}
  \bibinfo{author}{\bibfnamefont{F.}~\bibnamefont{Sette}},
  \bibinfo{journal}{Phys.\ Rev.\ Lett.} \textbf{\bibinfo{volume}{85}},
  \bibinfo{pages}{4076} (\bibinfo{year}{2000}{\natexlab{b}}).

\bibitem[{\citenamefont{Scopigno
  et~al.}(2000{\natexlab{c}})\citenamefont{Scopigno, Balucani, Ruocco, and
  Sette}}]{SBRSa}
\bibinfo{author}{\bibfnamefont{T.}~\bibnamefont{Scopigno}},
  \bibinfo{author}{\bibfnamefont{U.}~\bibnamefont{Balucani}},
  \bibinfo{author}{\bibfnamefont{G.}~\bibnamefont{Ruocco}}, \bibnamefont{and}
  \bibinfo{author}{\bibfnamefont{F.}~\bibnamefont{Sette}},
  \bibinfo{journal}{J.\ Phys.:\ Condens. Matter} \textbf{\bibinfo{volume}{12}},
  \bibinfo{pages}{8009} (\bibinfo{year}{2000}{\natexlab{c}}).

\bibitem[{\citenamefont{Wallace and Clements}(1999)}]{WCPRE59}
\bibinfo{author}{\bibfnamefont{D.~C.} \bibnamefont{Wallace}} \bibnamefont{and}
  \bibinfo{author}{\bibfnamefont{B.~E.} \bibnamefont{Clements}},
  \bibinfo{journal}{Phys.\ Rev.\ E} \textbf{\bibinfo{volume}{59}},
  \bibinfo{pages}{2942} (\bibinfo{year}{1999}).

\bibitem[{\citenamefont{Clements and Wallace}(1999)}]{CWPRE59}
\bibinfo{author}{\bibfnamefont{B.~E.} \bibnamefont{Clements}} \bibnamefont{and}
  \bibinfo{author}{\bibfnamefont{D.~C.} \bibnamefont{Wallace}},
  \bibinfo{journal}{Phys.\ Rev.\ E} \textbf{\bibinfo{volume}{59}},
  \bibinfo{pages}{2955} (\bibinfo{year}{1999}).

\bibitem[{\citenamefont{Wallace}(2002)}]{DWbook2}
\bibinfo{author}{\bibfnamefont{D.~C.} \bibnamefont{Wallace}},
  \emph{\bibinfo{title}{Statistical Physics of Crystals and Liquids}}
  (\bibinfo{publisher}{World Scientific, New Jersey}, \bibinfo{year}{2002}).

\bibitem[{\citenamefont{Wallace et~al.}(2001)\citenamefont{Wallace, Chisolm,
  and Clements}}]{WChCPRE64}
\bibinfo{author}{\bibfnamefont{D.~C.} \bibnamefont{Wallace}},
  \bibinfo{author}{\bibfnamefont{E.~D.} \bibnamefont{Chisolm}},
  \bibnamefont{and} \bibinfo{author}{\bibfnamefont{B.~E.}
  \bibnamefont{Clements}}, \bibinfo{journal}{Phys.\ Rev.\ E}
  \textbf{\bibinfo{volume}{64}}, \bibinfo{pages}{011205}
  (\bibinfo{year}{2001}).

\bibitem[{\citenamefont{Wallace et~al.}()\citenamefont{Wallace,
  {De~Lorenzi-Venneri}, and Chisolm}}]{WDCh05}
\bibinfo{author}{\bibfnamefont{D.~C.} \bibnamefont{Wallace}},
  \bibinfo{author}{\bibfnamefont{G.}~\bibnamefont{{De~Lorenzi-Venneri}}},
  \bibnamefont{and} \bibinfo{author}{\bibfnamefont{E.~D.}
  \bibnamefont{Chisolm}}, \bibinfo{howpublished}{to be published}.

\bibitem[{\citenamefont{Lovesey}(1984)}]{Lovbook}
\bibinfo{author}{\bibfnamefont{S.~W.} \bibnamefont{Lovesey}},
  \emph{\bibinfo{title}{Theory of Neutron Scattering from Condensed Matter}},
  vol.~\bibinfo{volume}{1} (\bibinfo{publisher}{Clarendon Press, Oxford},
  \bibinfo{year}{1984}).

\bibitem[{\citenamefont{Glyde}(1994)}]{Glybook}
\bibinfo{author}{\bibfnamefont{H.~R.} \bibnamefont{Glyde}},
  \emph{\bibinfo{title}{Excitations in Liquid and Solid Helium}}
  (\bibinfo{publisher}{Clarendon Press, Oxford}, \bibinfo{year}{1994}).

\bibitem[{\citenamefont{Glyde et~al.}(1977)\citenamefont{Glyde, Hansen, and
  Klein}}]{GHKPRB16}
\bibinfo{author}{\bibfnamefont{H.~R.} \bibnamefont{Glyde}},
  \bibinfo{author}{\bibfnamefont{J.~P.} \bibnamefont{Hansen}},
  \bibnamefont{and} \bibinfo{author}{\bibfnamefont{M.~L.} \bibnamefont{Klein}},
  \bibinfo{journal}{Phys.\ Rev.\ B} \textbf{\bibinfo{volume}{16}},
  \bibinfo{pages}{3476} (\bibinfo{year}{1977}).

\bibitem[{\citenamefont{Mazzacurati et~al.}(1996)\citenamefont{Mazzacurati,
  Ruocco, and Sampoli}}]{Mazz96}
\bibinfo{author}{\bibfnamefont{V.}~\bibnamefont{Mazzacurati}},
  \bibinfo{author}{\bibfnamefont{G.}~\bibnamefont{Ruocco}}, \bibnamefont{and}
  \bibinfo{author}{\bibfnamefont{M.}~\bibnamefont{Sampoli}},
  \bibinfo{journal}{Europhys.\ Lett.} \textbf{\bibinfo{volume}{34}},
  \bibinfo{pages}{681} (\bibinfo{year}{1996}).

\bibitem[{\citenamefont{Ruocco et~al.}(2000)\citenamefont{Ruocco, Sette,
  {Di~Leonardo}, Monaco, Sampoli, Scopigno, and Viliani}}]{R&cPRL00}
\bibinfo{author}{\bibfnamefont{G.}~\bibnamefont{Ruocco}},
  \bibinfo{author}{\bibfnamefont{F.}~\bibnamefont{Sette}},
  \bibinfo{author}{\bibfnamefont{R.}~\bibnamefont{{Di~Leonardo}}},
  \bibinfo{author}{\bibfnamefont{G.}~\bibnamefont{Monaco}},
  \bibinfo{author}{\bibfnamefont{M.}~\bibnamefont{Sampoli}},
  \bibinfo{author}{\bibfnamefont{T.}~\bibnamefont{Scopigno}}, \bibnamefont{and}
  \bibinfo{author}{\bibfnamefont{G.}~\bibnamefont{Viliani}},
  \bibinfo{journal}{Phys.\ Rev.\ Lett.} \textbf{\bibinfo{volume}{84}},
  \bibinfo{pages}{5788} (\bibinfo{year}{2000}).

\bibitem[{\citenamefont{Zwanzig}(1983)}]{ZJCP79}
\bibinfo{author}{\bibfnamefont{R.}~\bibnamefont{Zwanzig}},
  \bibinfo{journal}{J.\ Chem.\ Phys.} \textbf{\bibinfo{volume}{79}},
  \bibinfo{pages}{4507} (\bibinfo{year}{1983}).

\bibitem[{\citenamefont{{De~Lorenzi-Venneri} and Wallace}()}]{DW05}
\bibinfo{author}{\bibfnamefont{G.}~\bibnamefont{{De~Lorenzi-Venneri}}}
  \bibnamefont{and} \bibinfo{author}{\bibfnamefont{D.~C.}
  \bibnamefont{Wallace}}, \bibinfo{howpublished}{to be published}.

\bibitem[{\citenamefont{Wallace}(1998)}]{DWPRE58}
\bibinfo{author}{\bibfnamefont{D.~C.} \bibnamefont{Wallace}},
  \bibinfo{journal}{Phys.\ Rev.\ E} \textbf{\bibinfo{volume}{58}},
  \bibinfo{pages}{538} (\bibinfo{year}{1998}).

\bibitem[{\citenamefont{Chisolm et~al.}(2001)\citenamefont{Chisolm, Clements,
  and Wallace}}]{ChCWPRE63}
\bibinfo{author}{\bibfnamefont{E.~D.} \bibnamefont{Chisolm}},
  \bibinfo{author}{\bibfnamefont{B.~E.} \bibnamefont{Clements}},
  \bibnamefont{and} \bibinfo{author}{\bibfnamefont{D.~C.}
  \bibnamefont{Wallace}}, \bibinfo{journal}{Phys.\ Rev.\ E}
  \textbf{\bibinfo{volume}{63}}, \bibinfo{pages}{031204}
  (\bibinfo{year}{2001}).

\bibitem[{\citenamefont{Sciortino and Tartaglia}(1997)}]{ST97}
\bibinfo{author}{\bibfnamefont{F.}~\bibnamefont{Sciortino}} \bibnamefont{and}
  \bibinfo{author}{\bibfnamefont{P.}~\bibnamefont{Tartaglia}},
  \bibinfo{journal}{Phys.\ Rev.\ Lett.} \textbf{\bibinfo{volume}{78}},
  \bibinfo{pages}{2385} (\bibinfo{year}{1997}).

\end{thebibliography}

\end{document}